\documentclass[aps,pra,superscriptaddress,preprint,amsmath,amssymb,showpacs]{revtex4-2}
\usepackage[english]{babel}
\usepackage[utf8]{inputenc}
\usepackage{amssymb,bm}
\usepackage{graphicx}
\usepackage{amsmath}
\usepackage{color}
\usepackage{comment}
\usepackage[colorlinks=true,citecolor=blue,urlcolor=blue]{hyperref}
\begin{document}
 \begin{titlepage}

\title{Finite length of the solenoid and the Aharonov-Bohm effect}
\author{A.I.~Milstein}\email{a.i.milstein@inp.nsk.su}
\affiliation{Budker Institute of Nuclear Physics of SB RAS, 630090 Novosibirsk, Russia}
\affiliation{Novosibirsk State University, 630090 Novosibirsk, Russia}
\author{I.S.~Terekhov}\email{i.s.terekhov@gmail.com}
\affiliation{School of Physics and Engineering, ITMO University, 197101 St. Petersburg, Russia}

\date{\today}
\begin{abstract}
The scattering of a nonrelativistic electron on a narrow solenoid of finite length is considered. In this case, the magnetic field outside the solenoid is not zero. Using the eikonal approximation, the differential and total cross sections of the process are found. It is shown that the total cross section is finite, in contrast to the case of scattering on an infinitely long solenoid (Aharonov-Bohm effect). An asymmetry in the scattering cross section is also found, which can be observed in an experiment.	
\end{abstract}

\maketitle
 \end{titlepage}
\section{Introduction}

The scattering of an electron by a narrow infinitely long solenoid with a direct current, the Aharonov-Bohm effect \cite{AB1959}, is a purely quantum phenomenon. This is due to the fact that in the region outside the solenoid accessible to the motion of electron, the magnetic field is zero. Therefore, the Aharonov-Bohm effect has been studied in many papers (see, e.g., the review~\cite{PT1989}). Electron scattering has also been studied in the case of an infinitely long solenoid with an alternating current (the nonstationary Aharonov-Bohm effect \cite{LYG1992, ADC2000, GNS2011, SV2013, JJ2017, CM2019}). In the latter case, however, the magnetic and electric fields outside the solenoid are nonzero. Moreover, unlike the case of the standard Aharonov-Bohm effect, the total cross section of the process for the nonstationary Aharonov-Bohm effect is finite (see our recent work \cite{MT2025}, in which the differential and total scattering cross sections were calculated).

In the case of a narrow solenoid of finite length with direct current, the magnetic field outside the solenoid is nonzero. Therefore, it is of interest to study the effect of finite length on the differential and total scattering cross sections of the process. This is the goal of this work. The consideration is based on the use of the eikonal approximation \cite{LLQM}, applicable to our case. It show that the total cross section is finite. In addition, an asymmetry in the cross section arises, which can be observed experimentally.

\section{Eikonal approximation and the  cross section of process}
The vector potential $\bm A(\bm r)$ and the magnetic field $\bm B(\bm r)$ outside an infinitesimally thin solenoid of length $L$ are well known,
\begin{align}
	& \bm A(\bm r)=\dfrac{[\bm\nu\times\bm \rho ]}{\rho}A(\bm r)\,,\quad   A(\bm r)=\dfrac{\Phi}{4\pi \rho}\,\,\left\{\dfrac{z+L/2}{D_+(\bm r)}-\dfrac{z-L/2}{D_-(\bm r)}\right\},\nonumber\\
	&\bm B(\bm r)=  \dfrac{\Phi}{4\pi}\,\bm\nabla \left\{\dfrac{1}{D_+(\bm r)}-\dfrac{1}{D_-(\bm r)}\right\}\,,\nonumber\\
	&D_{\pm}(\bm r)=\sqrt{\rho^2+(z\pm L/2)^2}\,,
\end{align}
where $\bm\nu$ is a unit vector directed along the $z$-axis and parallel to the solenoid axis, $\bm r=(x,y,z)$, $\bm \rho=(x,y,0)$, $\rho=\sqrt{x^2+y^2}$, $\Phi$ is the magnetic flux through the solenoid, the origin is chosen at the center of the solenoid. From the experimental point of view, it is interesting to consider the motion of the electron beam perpendicular to $\bm\nu$ at $|z|\ll L$ and transverse width  $d\ll L$. In this region, the vector potential and magnetic field depend only on $\rho$,
\begin{align}
& A(\rho)=\dfrac{\Phi L}{4\pi \rho\,D(\rho)}\,,\quad
\bm B(\rho)= - \dfrac{\Phi L}{4\pi D^3(\rho)}\,\bm\nu \,,\nonumber\\
&D(\rho)=\sqrt{\rho^2+ L^2/4}\,.
\end{align}
Therefore, we can use the expressions  for the differential scattering cross section $d\sigma(\theta)$ and the total cross section $\sigma$ obtained in Ref.~\cite{MT2025} for the electron velocity $v$ and scattering angle $\theta$ satisfying the conditions $v/c\ll 1$ and $|\theta|\ll 1$, where $c$ is the speed of light. Recall that scattering occurs in a plane perpendicular to $\bm\nu$, the cross section has the dimension of length, and $-\pi\leq\theta\leq\pi$.

\begin{figure}[h!]
	\centering
	\includegraphics[width=0.46\linewidth]{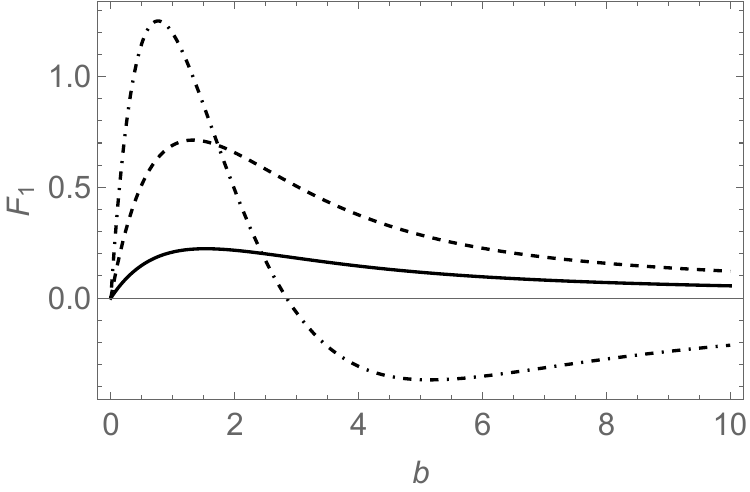}\hspace{1cm}\includegraphics[width=0.46\linewidth]{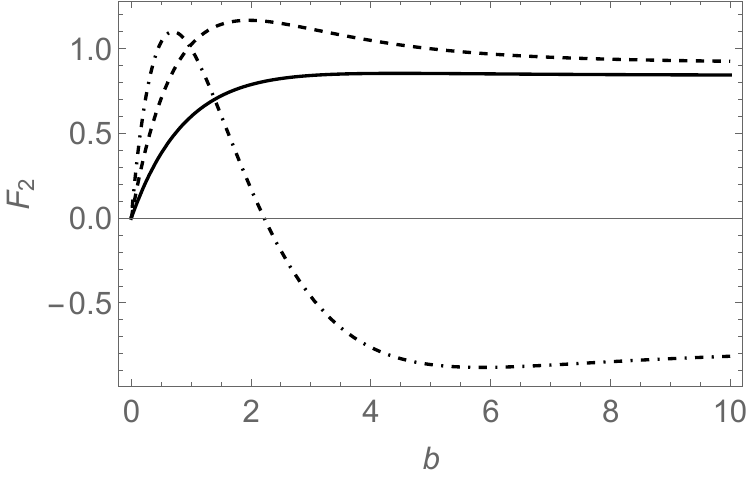}
	\caption{Dependence of functions $F_1(Q,b)$ (left plot) and $F_2(Q,b)$ (right plot) on $b$. Solid, dotted, and dash-dotted curves correspond to  $Q=1$,  $Q=2$, and $Q=4$, respectively.}
	\label{FigF1F2}
\end{figure}

Let the momentum of the incident electron be $P$ and directed along the $x$ axis. Then for the differential cross section we have \cite{MT2025}:
\begin{align} \label{secdif}
&d\sigma(\theta)=\dfrac{P}{2\pi\hbar}\,\left| \int_{-\infty}^{+\infty}dy\, e^{-iqy}\left[1-\,e^{-i\,\Xi(y)}\right]\,\right|^2d\theta\,,\nonumber\\
&\Xi(y)=\dfrac{ey}{c\hbar}\int_{-\infty}^\infty \dfrac{dx}{\rho}A(\rho)\,,
\end{align}	
where   $\hbar$ is the Planck constant, $q=P\theta/\hbar$, and $e$ is the electron charge. After elementary integration we find
\begin{align} \label{Xi}
&\Xi(y)=\dfrac{e\Phi}{\pi\hbar c}\,\mbox{sgn}(y)\, \arctan\left(\dfrac{L}{2|y|}\right)\,. \end{align}	
Substituting $\Xi(y)$ in the form \eqref{Xi} to Eq. \eqref{secdif} and integrating by parts over $y$,
we represent the differential scattering cross section as
\begin{align}\label{diffinal} 
	&d{\sigma}(\theta)=\dfrac{2\hbar}{\pi P}\,\left[F_1(Q,b)+sF_2(Q,b)\right]^2\dfrac{d\theta}{\theta^2}\,,\nonumber\\ &F_1(Q,b)=\dfrac{2Q}{\pi}\int_{0}^{\infty}\dfrac{dt}{1+t^2}\sin(bt)\sin\left[\dfrac{2Q}{\pi}\arctan\dfrac{1}{t}\right]\,,\nonumber\\
	&F_2(Q,b)=\sin Q-\dfrac{2Q}{\pi}\int_{0}^{\infty}\dfrac{dt}{1+t^2}\cos(bt)\cos\left[\dfrac{2Q}{\pi}\arctan\dfrac{1}{t}\right]\,,\nonumber\\
	& s=\mbox{sgn}\,\theta\,,\quad b=\dfrac{|\theta|}{\theta_0}\,,\quad \theta_0=\dfrac{2\hbar}{LP}\,,\quad Q=\dfrac{e\Phi}{2 c\hbar}\,.
\end{align}	
The dependence of functions $F_1(Q,b)$ and $F_2(Q,b)$ on $b$ for a few values of $Q$ are shown in Fig.~\ref{FigF1F2}.

For $b\gg 1$, that is, $|\theta|\gg\theta_0$, the asymptotics of the functions $F_{1,2}(Q,b)$ are 
\begin{align} 
 &F_1(Q,b)=\dfrac{2Q\sin Q}{\pi b}\,,\quad F_2(Q,b)=\sin Q\,.
\end{align}	
As a result, at $|\theta|\gg\theta_0$ the cross section has the form
\begin{align}\label{sigtheta1} 
& d\sigma(\theta) =\dfrac{2\hbar}{\pi P}\,\sin^2Q\,\left(\dfrac{\theta_1}{|\theta|}+s\right)^2\,\dfrac{d\theta}{\theta^2}\,,\nonumber\\
&\theta_1=\dfrac{2Q}{\pi}\,\theta_0=\dfrac{2e\Phi}{\pi LcP} \,. 
\end{align}
For $|\theta|\gg\theta_1$ we obtain the known answer for the cross section in the case of  standard Aharonov-Bohm effect \cite{AB1959,LLQM},
\begin{align}\label{AB} 
& d\sigma(\theta) =\dfrac{2\hbar}{\pi P}\,\sin^2Q\,\dfrac{d\theta}{\theta^2}\,.
\end{align}
For a reasonable value of parameters $\theta_0$ is very small. However, for $Q\gg 1$ the value $\theta_1$ is not very small, since $\theta_1$ is independent of $\hbar$. So, for the solenoid of length $L=10\, \text{cm}$, magnetic flux $\Phi=10\, \text{G}\cdot\text{cm}^2$, and an electron having energy $10\, \text{eV}$ we obtain $\theta_0=1.3\times 10^{-9}$  and $\theta_1=0.13$. Note that the term $\theta_1/\theta$ in Eq.~\eqref{sigtheta1} reflects the effect of nonzero magnetic field outside the solenoid on the electron scattering cross section.

\begin{figure}[h!]
	\centering
	\includegraphics[width=0.5\linewidth]{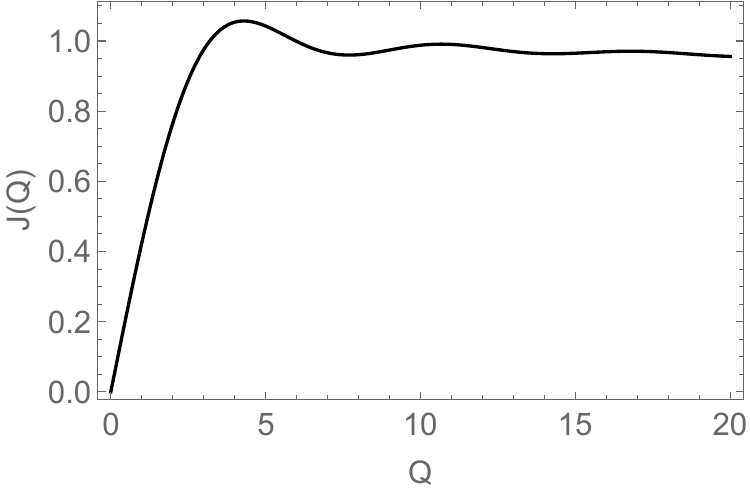}
		\caption{The function $J(Q)=\sigma/(2LQ)$, see Eq.~\eqref{eqJ}.  }
	\label{JQ}
\end{figure}

For $b\ll 1$, that is, $|\theta|\ll\theta_0$, the asymptotics read
\begin{align}\label{eqJ} 
&F_1(Q,b)=b\,Q\,J(Q)\,,\quad F_2(Q,b)=b\, Q\,,\nonumber\\
&J(Q)=\dfrac{2}{\pi}\int_{0}^{\infty}\dfrac{t\,dt}{1+t^2}\sin\left[\dfrac{2Q}{\pi}\arctan\dfrac{1}{t}\right]\,.
\end{align}	
The differential cross section in this region is 
\begin{align} 
d\sigma(\theta) =\dfrac{2\hbar\, Q^2}{\pi P\,\theta_0^2}\,(1+J^2+2sJ)\,d\theta\,. 
\end{align}
One can see, that the quantity $d\sigma(\theta)$ has no singularity at small $\theta$, therefore the total cross section is finite. To find the total cross section, we integrate over $\theta$ in Eq. \eqref{secdif} and obtain:
\begin{align} \label{sigtot}
&\sigma=4\int_0^\infty [1-\cos\,\Xi(y)]dy=2LQ\,J(Q)\,.
\end{align}	
The function $J(Q)$ is shown in Fig.~\ref{JQ}. At small and large values of the parameter $Q$ the function $J(q)$ has following behavior:
\begin{eqnarray}
	J(Q)=\left\{
	\begin{array}{cc}
	\frac{2\ln 2}{\pi}Q, & Q\ll1,\\
	1,& Q\gg 1.
	\end{array}\right.
\end{eqnarray}
 
It follows from the asymptotics of the functions $F_{1,2}(Q,b)$  that the differential cross section at $\theta\gg\theta_1$ is independent of $s=\mbox{sgn}\,\theta$.
At $\theta\sim \theta_1$, there is asymmetry  in the cross section due to the interference of the functions $F_1(Q,b)$ and $F_2(Q,b)$.
We represent the  $d\sigma(\vartheta)$ in Eq.~\eqref{diffinal} as
\begin{align} \label{difsa}
	&d\sigma(\theta)=d\sigma_{s}(\theta)+s\,d\sigma_{a}(\theta)\,,\nonumber\\
	&d{\sigma}_{s}(\theta)=\dfrac{2\hbar}{\pi P}\,\left[F_1^2(Q,b)+F_2^2(Q,b)\right]\dfrac{d\theta}{\theta^2}\,,\nonumber\\
	&d{\sigma}_{a}(\theta)=\dfrac{4\hbar}{\pi P}\,F_1(Q,b)F_2(Q,b)\dfrac{d\theta}{\theta^2}\,.
\end{align}	
We define the asymmetry $\Sigma_a$ as follows
\begin{align} \label{asymdef}
&\Sigma_a(\theta)=\dfrac{d\sigma_a}{d\sigma_s}=\dfrac{2F_1(Q,b)F_2(Q,b)}{F_1^2(Q,b)+F_2^2(Q,b)}\,. 
\end{align}	
At $|\theta|\gg\theta_0$, the asymmetry has the form 
\begin{align} \label{asymas1}
&\Sigma_a(\theta)=\dfrac{2\,\theta_1|\theta|}{\theta^2_1+\theta^2}\,,
\end{align}	
and tends to zero at $|\theta|\gg\theta_1$.  In principle, this asymmetry can be observed experimentally,   see the discussion after Eq. \eqref{AB}. 
At $|\theta|\ll\theta_0$ we have
\begin{align} \label{asymas2}
&\Sigma_a=\dfrac{2J}{1+J^2}\,. 
\end{align}	
This asymptotics is independent of $\theta$. It follows from Fig.~\ref{JQ} that the asymmetry  $\Sigma_a(\theta)$  at $|\theta|\lesssim \theta_1$ is not a small quantity. At $|\theta|\ll \theta_1$ and $Q\to\infty$ the asymmetry tends to unity.

\section{Conclusion}
In the nonrelativistic limit, the effect of a finite length $L$ of a narrow solenoid on the electron scattering cross section is investigated. Using the eikonal approximation, the differential and total scattering cross sections are found. In contrast to the standard Aharonov-Bohm effect corresponding to an infinite length $L$, the total cross section in our case is finite. The main contribution to $\sigma$ is given by the angles $|\theta|\sim \theta_0=2\hbar/LP\ll 1$. For a finite $L$, a new effect arises: the differential scattering cross section contains an asymmetry with respect to the replacement $\theta\to -\theta$. This asymmetry has a maximum at $|\theta|\sim \theta_1$ and rapidly decreases at $|\theta|\gg \theta_1$. This asymmetry is not a small quantity and can be observed experimentally.

\section*{Acknowledgement}
The work of Ivan Terekhov was financially supported by the ITMO Fellowship Program.

\end{document}